\documentclass[11pt]{article}

\usepackage{jcappub} 
\usepackage{mathrsfs}     
\usepackage{bm}     
\usepackage{hyperref} 
\hypersetup{
    colorlinks=true,       
    linkcolor=red,          
    citecolor=blue,        
    filecolor=magenta,      
    urlcolor=blue           
}
\usepackage[all]{hypcap} 
\usepackage{graphicx} 	
\usepackage{amsmath}
\usepackage{physics} 

\newcommand{\ba}[1]{\begin{align} #1 \end{align}}
\newcommand{\bes}[1]{\begin{equation}\begin{split} #1 \end{split}\end{equation}}
\newcommand{\bsa}[2]{\begin{subequations}\label{#1}\begin{align} #2 \end{align}\end{subequations}}

\newcommand{\fref}[1]{figure~\ref{#1}}
\newcommand{\Fref}[1]{Figure~\ref{#1}}

\newcommand{\sref}[1]{section~\ref{#1}}

\newcommand{\aref}[1]{appendix~\ref{#1}}

\newcommand{\eref}[1]{eq.~(\ref{#1})}
\newcommand{\erefs}[2]{eqs.~(\ref{#1})~and~(\ref{#2})}
\newcommand{\Eref}[1]{Eq.~(\ref{#1})}
\newcommand{\rref}[1]{ref.~\cite{#1}}
\newcommand{\rrefs}[1]{refs.~\cite{#1}}

\newcommand{\nhat}{\bm{n}}

\newcommand{\hatNloops}{\hat{N}_{\rm loops}}

\title{Measures of non-Gaussianity in axion-string-induced CMB birefringence}

\date{\today}

\author[a]{Ray Hagimoto}
\author[a]{and Andrew J. Long}

\affiliation[a]{Department of Physics and Astronomy, Rice University, \\ 6100 Main Street, Houston, TX 77005, U.S.A.}

\emailAdd{rmh14@rice.edu}
\emailAdd{andrewjlong@rice.edu}

\abstract{
The presence of axion strings in the Universe after recombination can leave an imprint on the polarization pattern of the cosmic microwave background radiation through the phenomenon of axion-string-induced birefringence via the hyperlight axion-like particle's coupling to electromagnetism.  Across the sky, the polarization rotation angle is expected to display a patchwork of uniform regions with sharp boundaries that arise as the `shadow' of axion string loops.  The statistics of such a birefringence sky map are therefore necessarily non-Gaussian.  In this article we quantify the non-Gaussianity in axion-string-induced birefringence using two techniques, kurtosis and bispectrum, which correspond to $4$- and $3$-point correlation functions. If anisotropic birefringence were detected in the future, a measurement of its non-Gaussian properties would facilitate a discrimination across different new physics sources generally, and in the context of axion strings specifically, it would help to break degeneracies between the axion-photon coupling and properties of the string network.  
}


\begin{document}
\maketitle
\flushbottom

\section{Introduction}
\label{sec:intro}

Observations of the cosmic microwave background (CMB) temperature and polarization anisotropies have informed our understanding of the composition, structure, and evolution of the Universe.  
These precision measurements have also revealed some surprises, such as the mysterious dark matter and dark energy that permeate the Universe.  
Ongoing and future observations, with significantly higher precision, may uncover evidence for additional cosmological relics that are currently out of reach \cite{CMB-HD:2022bsz,CMB-S4:2022ght} such as cosmic axion strings.  
In this work we seek to quantify the signatures of axion strings through their non-Gaussian cosmic birefringence. 

Cosmic strings, one-dimensional topological defects formed from scalar fields \cite{Vilenkin:2000jqa}, are predicted to arise in the early universe during phase transitions associated with as-yet undiscovered new physics.  
While the new particles and forces may be inaccessible, because they are too heavy to be produced at high-energy colliders or too feebly coupled to be probed in the laboratory, the network of cosmic strings can leave a detectable imprint on the CMB radiation, which is both exquisitely measured and theoretically well understood.  
For example, searches for the gravitational influence of cosmic strings on the CMB anisotropies have already yielded an upper limit on the strings' tension \cite{Charnock:2016nzm,Lizarraga:2016onn}, which translates into a strong constraint on the scale of new physics.
On the other hand, if the string-forming fields couple non-gravitationally to visible matter and radiation, novel channels for testing these theories become available.  
Cosmic strings formed from hyperlight axion-like particles (ALPs) that couple to electromagnetism provide an especially compelling target, since they are expected to induce a birefringence of CMB polarization \cite{Agrawal:2019lkr}. 

The phenomenon of axion-induced birefringence has been a subject of great interest for many years \cite{Harvey:1988in,Carroll:1989vb,Carroll:1991zs,Harari:1992ea,Carroll:1998bd,Fedderke:2019ajk,Greco:2022ufo,Murai:2022zur,Cai:2022zad,Greco:2022xwj,Yin:2023srb,Naokawa:2023upt}.  
The important aspect of birefringence from axion strings \cite{Agrawal:2019lkr} is that the typical axion field excursion is large $\Delta a \approx 2 \pi f_a$ , thereby evading a suppression factor that appears for other models, such as axion dark matter.  
Several recent studies \cite{Agrawal:2020euj,Takahashi:2020tqv,Jain:2021shf,Yin:2021kmx,Kitajima:2022jzz,Jain:2022jrp,Gonzalez:2022mcx,Yin:2023vit} have explored the signatures of axion-string-(and domain wall)-induced birefringence, calculated the angular power spectrum, and assessed compatibility with the various measurements of CMB birefringence (including a claimed detection of isotropic birefringence \cite{Minami:2020odp,Diego-Palazuelos:2022dsq,Eskilt:2022wav,Eskilt:2022cff,Diego-Palazuelos:2022cnh,Eskilt:2023ndm}).  
To summarize, these studies conclude that the current generation of CMB telescopes (Planck, SPT{\sc pol}, ACT{\sc pol}, BICEP2/\textit{Keck Array}, {\sc Polarbear}) are nearly sensitive enough to probe the most well-motivated parameter space, and next-generation telescopes will put these theories to the test.  

\begin{figure}[t]
      \centering
      \includegraphics[width=0.8\textwidth]{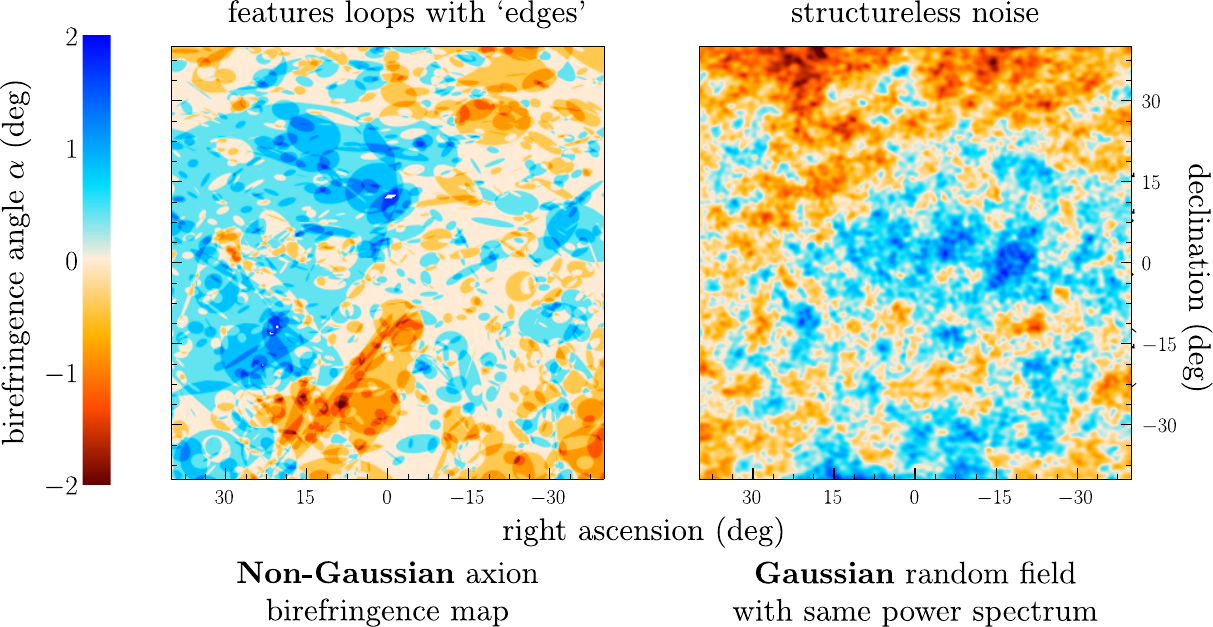}
      \caption{\label{fig:maps}
      \textit{Left:} a simulated map of the birefringence angle $\alpha(\hat{\bm n})$ for an axion string network.  \textit{Right:} a simulated map assuming Gaussian statistics with the same power spectrum as the left map.  The disk-like features on the left map are a manifestation of the non-Gaussian nature of the stochastic variable.  The non-Gaussian map is generated using the loop-crossing model with $\zeta_0 = \xi_0 = \mathcal{A} = 1$; see \sref{sec:kurtosis} for additional details. 
      }
\end{figure}

Whereas most of the work on axion-string-induced birefringence has focused thus far on two-point statistics such as the angular power spectrum, the higher moments contain a wealth of valuable information that could help to discriminate across different sources of birefrigence \cite{Yin:2023vit} if a detection were made with next-generation surveys \cite{Guzman:2021ygf}. 
We illustrate this point in \fref{fig:maps}; the left panel shows a simulated map of the birefringence angle across a patch of the sky arising from a network of axion strings, and the right panel shows a map that was simulated using Gaussian statistics with the same angular power spectrum. 
These two images can be distinguished easily: the map on the left displays disk-like structures, corresponding to the imprint of axion string loops.  
Since these birefringence maps have the same two-point correlations, the difference between them arises from higher-order correlations, which cannot reduce to two-point correlations for non-Gaussian statistics.  

In this work we seek to quantify these non-Gaussian features in axion-string-induced birefringence using three-point correlations (bispectrum) and four-point correlations (kurtosis), which are familiar tools from studies of CMB non-Gaussianity \cite{Meerburg:2019qqi}.  
Similar techniques have been used in the past \cite{Gangui:1996xxx,Landriau:2010cb,Planck:2013mgr,Hergt:2016xup} to search for evidence of a cosmic string network's gravitational influence of the CMB anisotropies.  
Our approach is complementary to the one taken in \rref{Yin:2023vit}, which contains a related analysis of axion-string-induced birefringence using the scattering transform.  

\section{Kurtosis}
\label{sec:kurtosis}

We denote the birefringence map by $\hat{\alpha}({\bm n})$ where ${\bm n}$ is a unit vector, indicating a direction on the sky, and $\hat{\alpha}$ is the birefringence angle, corresponding to the rotation of the polarization axis.  
We use hats to denote random variables and angled brackets to denote ensemble averaging.  
The birefringence map admits a multipole expansion: 
\ba{
    \hat{\alpha}({\bm n}) = \sum_{\ell=0}^{\infty} \sum_{m=-\ell}^{\ell} \hat{\alpha}_{\ell m} Y_{\ell m}({\bm n}) 
}
where $\hat{\alpha}_{\ell m}$ are called the multipole moment coefficients and $Y_{\ell m}({\bm n})$ are the spherical harmonics; we use the standard normalization $\int \! \mathrm{d}^2 {\bm n} \, |Y_{\ell m}({\bm n})|^2 = 1$.  
Since the birefringence map is real $\hat{\alpha}({\bm n})^\ast = \hat{\alpha}({\bm n})$, the complex multipole moment coefficients obey $\hat{\alpha}_{\ell m}^\ast = (-1)^m \, \hat{\alpha}_{\ell -m}$.  

Kurtosis is a convenient measure of non-Gaussianity that is both easy to calculate and intuitive to understand.  
The kurtosis of the (complex) multipole moment coefficients is given by 
\ba{\label{eq:kurtosis}
    \kappa_{\ell m} 
    = \frac{\bigl< \bigl| \hat{\alpha}_{\ell m} - \langle \hat{\alpha}_{\ell m} \rangle \bigr|^4 \bigr>}{\bigl< \bigl| \hat{\alpha}_{\ell m} - \langle \hat{\alpha}_{\ell m} \rangle \bigr|^2 \bigr>^2} 
    = \frac{\bigl< \bigl| \hat{\alpha}_{\ell m} \bigr|^4 \bigr>}{\bigl< \bigl| \hat{\alpha}_{\ell m} \bigr|^2 \bigr>^2} 
    \;, 
}
where the first equality is the general definition, and the second equality holds for axion-string-induced birefringence that has vanishing 1-point functions $\langle \hat{\alpha}_{\ell m} \rangle = 0$.  
If the real and imaginary parts of the multipole moment coefficients were $\mathrm{i.i.d.}$ Gaussian random variables, then Isserlis's theorem (Wick's theorem) would reduce the 4-point functions to products of 2-point functions.  
For modes with $m=0$ the reality condition forces $\hat{\alpha}_{\ell 0}$ to be real implying $\kappa_{\ell m} = 3$, whereas for $m \neq 0$ the complex $\hat{\alpha}_{\ell m}$ would have $\kappa_{\ell m} = 2$ instead.\footnote{
For a single Gaussian random variable $\hat{x}$ with $\langle \hat{x} \rangle = 0$, one finds $\langle \hat{x}^4 \rangle = 3 \langle \hat{x}^2 \rangle^2$ and the kurtosis is $\langle \hat{x}^4 \rangle / \langle \hat{x}^2 \rangle^2 = 3$.  For a complex random variable $\hat{X} = \hat{x} + i \hat{y}$ with statistically independent real and imaginary parts $\langle \hat{x} \hat{y} \rangle = 0$, one finds instead $\langle |\hat{X}|^4 \rangle = \langle (\hat{x}^2 + \hat{y}^2)^2 \rangle = 3 \langle \hat{x}^2 \rangle^2 + 2 \langle \hat{x}^2 \rangle \langle \hat{y}^2 \rangle + 3 \langle \hat{y}^2 \rangle^2$ and $\langle |\hat{X}|^2 \rangle^2 = \langle \hat{x}^2 \rangle^2 + 2 \langle \hat{x}^2 \rangle \langle \hat{y}^2 \rangle + \langle \hat{y}^2 \rangle^2$, and the kurtosis is $\langle |\hat{X}|^4 \rangle / \langle |\hat{X}|^2 \rangle^2 = 2$ for $\langle \hat{x}^2 \rangle = \langle \hat{y}^2 \rangle$.
}
We define the excess kurtosis
\ba{\label{eq:excess-kurtosis}
    \Delta \kappa_{\ell m} = \begin{cases} 
    \kappa_{\ell 0} - 3 & , \quad \text{for $m = 0$} \\ 
    \kappa_{\ell m} - 2 & , \quad \text{for $m \neq 0$} \\ 
    \end{cases}
    \;,
}
which vanishes for Gaussian statistics.  
A positive excess kurtosis $\Delta \kappa_{\ell m} > 0$ corresponds to a distribution with a tighter center and broader tails than a Gaussian having the same mean and variance.  
In this way, kurtosis provides an intuitive measure of the departure from Gaussianity.  

We seek to employ kurtosis as a measure of non-Gaussianity in axion-string-induced birefringence maps.  
To that end, we simulate birefringence maps using the loop-crossing model (LCM), as described in \rrefs{Jain:2021shf,Jain:2022jrp}.  
The LCM is informed by simulations of axion string networks including refs.~\cite{Yamaguchi:1998gx,Yamaguchi:2002sh,Hiramatsu:2010yu,Hiramatsu:2012gg,Kawasaki:2014sqa,Lopez-Eiguren:2017dmc,Gorghetto:2018myk,Kawasaki:2018bzv,Vaquero:2018tib,Martins:2018dqg,Buschmann:2019icd,Hindmarsh:2019csc,Klaer:2019fxc,Gorghetto:2020qws,Gorghetto:2021fsn,Hindmarsh:2021vih,Buschmann:2021sdq}.  
In this model, the string network is approximated as a collection of circular planar loops with a statistically homogeneous distribution through space and a statistically isotropic orientation.  
On cosmological time scales, the number density of loops decreases and the length of loops grows so as to track the cosmological expansion.  
Specifically, the number density of loops at time $t$ is $n(t) = \xi_0 H(t)^3 / 2 \pi \zeta_0$ and the radius of loops at time $t$ is $\zeta_0 / H(t)$ where $H(t)$ is the Hubble parameter.
The dimensionless coefficients, $\xi_0$ and $\zeta_0$, are two model parameters, and string network simulations motivate values around $\xi_0 = 1$-$10$ and $\zeta_0 = 0.1$-$1$.  
As a photon propagates through the string network, from the CMB to a detector on Earth, birefringence accumulates each time the photon passes through the disk bounded by a string loop.  
The birefringence induced by each loop crossing is $\pm \mathcal{A} \alpha_\mathrm{em}$ where the dimensionless anomaly coefficient $\mathcal{A} = 0.1$-$1$ is another model parameter, $\alpha_\mathrm{em} \simeq 1/137$ is the electromagnetic fine structure constant, and the two equally-probable signs $\pm 1$ depend on the relative orientation of the loop and the photon's propagation direction.  
The signal of axion-string-induced birefringence also depends upon the axion mass scale $m_a$, since the string network forms domain walls when the Hubble parameter is comparable to the axion mass scale, possibly suppressing the signal of axion-string-induced birefringence \cite{Jain:2022jrp}.  
In this work, we assume that the axion mass is smaller than the Hubble scale today $m_a \lesssim 3 H_0$ and the string network survives until present times, allowing for an unsuppressed birefringence signal.  
We implement the loop-crossing model in a Python code that interfaces with HEALPix \cite{2005ApJ...622..759G,Zonca2019} taking $N_\mathrm{side} = 128$ or $512$ for different studies in this work.  
With a large number of simulated birefringence maps we calculate sample means to estimate ensemble averages and thereby evaluate the kurtosis of the multipole moment coefficients.  

\begin{figure}[t!]
      \centering
      \includegraphics[width=0.62\textwidth]{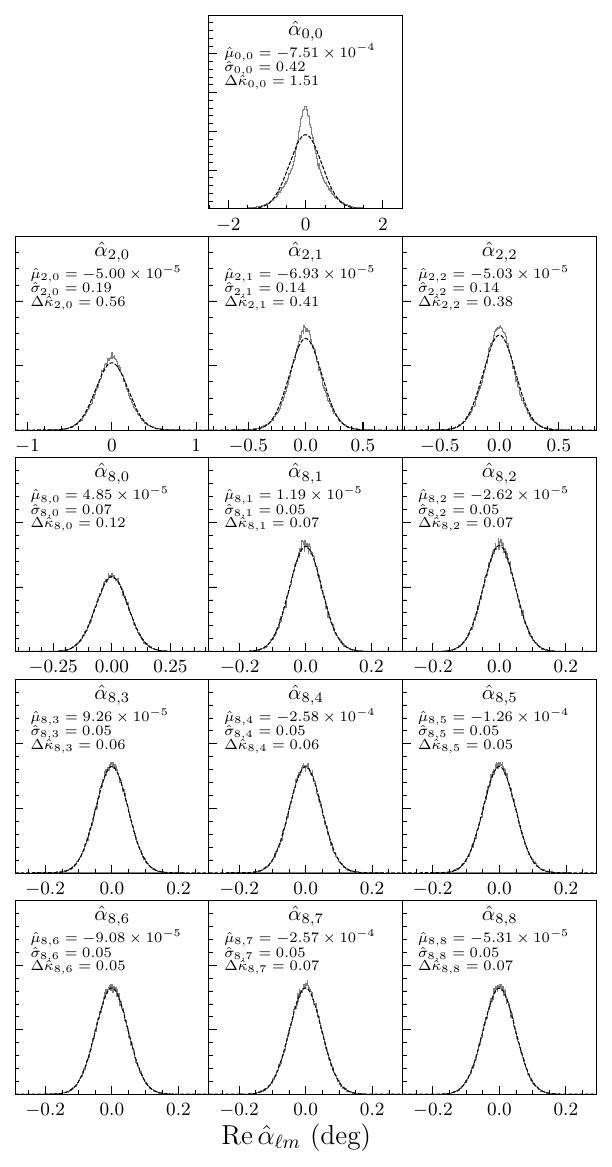}
      \caption{\label{fig:alpha_lm_distributions}
      Histogram showing distributions over the real part of the multipole moment coefficients $\hat{\alpha}_{\ell m}$ for axion-string-induced birefringence. These distributions were generated using $150,\!000$ realizations of birefringence maps simulated in the loop-crossing model with $\zeta_0 = 1$, $\xi_0 = 1$, and $\mathcal{A} = 1$. In each panel the $x$-axis is the value of $\Re \hat{\alpha}_{lm}$ in degrees, $\hat{\mu}_{\ell m}$ and $\hat{\sigma}_{\ell m}$ are the sample mean and standard deviation of $\Re \hat{\alpha}_{\ell m}$, and $\Delta \hat{\kappa}_{\ell m}$ is the excess kurtosis of $\hat{\alpha}_{\ell m}$ through \erefs{eq:kurtosis}{eq:excess-kurtosis}. Black dashed curves show a Gaussian distribution with the same mean and variance as the histogram.
      }
\end{figure}

Using the loop-crossing model, we obtain more than $60,\!000$ simulated realizations (up to $150,\!000$) of the axion-string-induced birefringence map. 
For each map we extract the multipole moment coefficients $\hat{\alpha}_{\ell m}$.  
To assess the departure from Gaussianity, we show in \fref{fig:alpha_lm_distributions} the distributions over $\Re \hat{\alpha}_{\ell m}$ for the lowest several multipole moments.  
The distributions over $\Im \hat{\alpha}_{\ell m}$ (not shown) are similar.  
We only show multipole moments with $m > 0$ since the reality condition imposes $\Re \hat{\alpha}_{\ell m} = (-1)^{m} \, \Re \hat{\alpha}_{\ell -m}$.  
We give values of the sample mean $\hat{\mu}_{\ell m}$, sample standard deviation $\hat{\sigma}_{\ell m}$, and sample excess kurtosis $\Delta \hat{\kappa}_{\ell m}$ that were inferred from the suite of simulations.  
To highlight the departure of these distributions from Gaussianity, we show a normal distribution (dashed line) with the same mean and variance as each histogram. 
The histograms are approximately symmetric and centered close to zero, since each loop crossing shifts the birefringence by $\pm \mathcal{A} \alpha_\mathrm{em}$ with equal probability.  

\Fref{fig:alpha_lm_distributions} displays a departure from Gaussianity for multipole moments with small index $\ell$.  
For $\ell = 0$ and $2$, the distinction between the histogram and the normal distribution is clearly evident.  
One can easily see that the histogram is tighter and taller around $\hat{\alpha}_{\ell m} = 0$, and close inspection reveals that it also has wider tails.  
In general such features correspond to a positive excess kurtosis.  
For the monopole we find the excess kurtosis to be $\Delta \hat{\kappa}_{0,0} \approx 1.51$; for the quadrupole it is $\Delta \hat{\kappa}_{2,m} \approx 0.4$; and for $\ell = 8$ is it $\Delta \hat{\kappa}_{8,m} \approx 0.06$. 
For a given $\ell$ we find that each $m$ has a similar distribution, which is consistent with the underlying statistical isotropy of the loop-crossing model.  
These examples illustrate that the excess kurtosis decreases as the multipole index $\ell$ increases.  

\begin{figure}[t!]
\centering
    \includegraphics{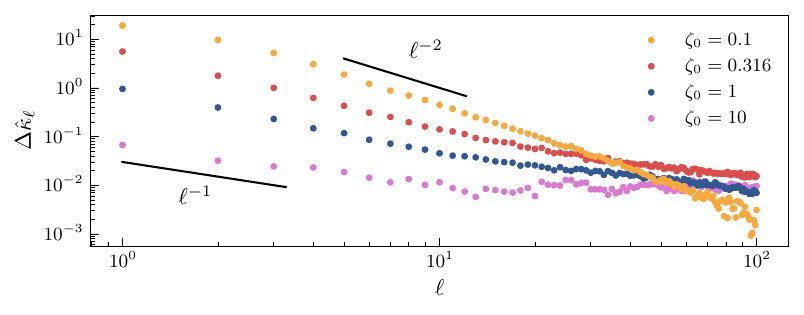}
    \caption{\label{fig:kurtosis}
    Excess kurtosis of axion-string-induced birefringence for a range of multipole moments.  We show the average excess kurtosis $\Delta \hat{\kappa}_\ell$ for multipole moments with index $\ell$ ranging from $1$ to $100$. The excess kurtosis is calculated from simulated birefringence maps that were created using the loop-crossing model with $\xi_0 = 1$, $\mathcal{A} = 1$, and three values of the loop-length parameter $\zeta_0$. The number of realizations is $150,\!000$ for $\zeta_0 = 1$, $40,\!000$ for $\zeta_0 = 0.316$, $62,\!000$ for $\zeta_0 = 0.1$, and $150,\!000$ for $\zeta_0 = 10$. The curves approximately follow broken power law scalings for small and large multipoles. 
    }
\end{figure}

We are interested in how the kurtosis varies across angular scales, and specifically how quickly the excess kurtosis decreases for higher multipole moments.  
Since the loop-crossing model generates a statistically isotropic birefringence map, we expect that $\Delta \hat{\kappa}_{\ell m}$ should only depend on the index $\ell$. 
This observation motivates us to define the `angle-averaged' excess kurtosis 
\ba{
    \Delta \hat{\kappa}_\ell & \equiv 
    \frac{1}{\ell} \sum_{m = 1}^{\ell} \Delta \hat{\kappa}_{\ell m} 
    \;,
}
for $\ell > 0$.  
In \fref{fig:kurtosis} we show the average excess kurtosis across a range of angular scales corresponding to multipole moment indices $\ell = 1$ to $100$.  
We obtain these numerical results from simulated birefringence maps obtained through the loop-crossing model with four values of the dimensionless loop-length parameter: $\zeta_0 = 10$ corresponding to ten-times Hubble scale loops, $\zeta_0 = 1$ corresponding to Hubble scale loops, $\zeta_0 = 10^{-0.5} \approx 0.316$, and $\zeta_0 = 0.1$ corresponding to loops that are a tenth of the Hubble scale.  
Kurtosis is independent of the parameter $\mathcal{A}$, since 
$\hat{\alpha}_{\ell m} \propto \mathcal{A}$ and this factor cancels when calculating kurtosis as a ratio of multipole moment coefficients through \eref{eq:kurtosis}.  

\Fref{fig:kurtosis} exhibits several notable features.  
We observe that:  
(1) the excess kurtosis is positive across this range of multipoles and for this set of model parameters; 
(2) the excess kurtosis is $\approx 0.1$-$10$ at low multipoles, and its value goes inversely with $\zeta_0$; and 
(3) the excess kurtosis decreases with increasing multipole index $\ell$ in a (statistically) monotonic way, and it approximately follows a broken power law scaling.  
These features can be loosely understood as follows.  
For $\zeta_0 = 1$ the string network contains only one or two loops at the largest angular scales (smallest $\ell$), and the statistics of the birefringence map should also be order one numbers implying $\hat{\kappa}_{\ell m} \approx 1$ and $\Delta \hat{\kappa}_{\ell} \approx 1$.  
Increasing $\ell$ corresponds to decreasing the angular scale of interest, and the network contains more loops at smaller scales on average.  
As the number of loops increases, their imprint on the birefringence map corresponds to many overlapping disks and ellipses that each contribute $\pm \mathcal{A} \alpha_\mathrm{em}$.  
Since each loop's contribution can be either positive or negative (with equal probability, associated with the random orientation of the loop), the net birefringence grows like a random walk with a random number of steps.  
By the central limit theorem, the statistics of this quantity converge to Gaussian in the limit of many loops.  
Consequently, one expects an increasingly Gaussian birefringence map on smaller angular scales, corresponding to an inverse relationship between $\ell$ and $\Delta \hat{\kappa}_\ell$, such as the one seen in \fref{fig:kurtosis}. 
Furthermore, one expects the excess kurtosis to be positive, because the non-Gaussianity is primarily driven by the fact that there are few large loops. 
These rare outliers boost the tails of the $\hat{\alpha}_{\ell m}$ distribution at values that are relatively large compared to the standard deviation; such features are characteristic of a distribution with positive excess kurtosis.

The preceding loose argument can be formulated more concretely for the monopole $\hat{\alpha}_{00}$, which is proportional to the sky-average birefringence angle.  
This analysis is presented in \aref{app:kurtosis}.  
We find that the excess kurtosis in the monopole is inversely proportional to the average number of loops, $\Delta \hat{\kappa}_0 = 1 / \bar{N}_\mathrm{loops}$.  
Extending this scaling to the higher multipole moments suggests the relationship $\Delta \hat{\kappa}_\ell \sim 1 / \bar{N}_\ell$, where $\bar{N}_\ell$ is the average number of loops at a given angular scale $\sim \pi/\ell$.  
By evaluating the average number of loops as a function of $\ell$ and the string network model parameters, $\zeta_0$ and $\xi_0$, we obtain an analytical estimate of the excess kurtosis 
\begin{align}\label{eq:analytic-kurtosis-vs-multipole}
    \Delta \hat{\kappa}_\ell
    \sim 
    \frac{\zeta_0}{8 \xi_0}\,
    \biggl( 1 + \frac{\pi}{\lambda \zeta_0 \ell} \biggr)^2
    \;.
\end{align}
Here $\lambda \approx 0.3$ is a constant numerical factor.  
See \aref{app:kurtosis} for the derivation of \eref{eq:analytic-kurtosis-vs-multipole}.  

The analytical formula in \eref{eq:analytic-kurtosis-vs-multipole} agrees well with the numerical results presented in \fref{fig:kurtosis}.  
\Eref{eq:analytic-kurtosis-vs-multipole} implies that $\Delta \hat{\kappa}_\ell$ should scale like $\ell^{-2}$ for $\ell \ll \pi / (\lambda\,\zeta_0) \approx 10 / \zeta_0$ and like $\ell^0$ for larger $\ell$.  
Similarly, \fref{fig:kurtosis} shows an $\ell^{-2}$ scaling for small values of $\ell$, and a flattening (in the $\zeta_0 = 10$, $1$, and $0.316$ curves) for larger values of $\ell$ approaching $\ell = 100$.  
Additionally, the angular scale dividing these two regimes is well approximated by $10 / \zeta_0$. 
For the $\zeta_0 = 0.1$ curve, the flattening is not seen, and this is compatible with the analytical model since the transition scale $10 / \zeta_0 \approx 100$, and the full plot range from $\ell = 1$ to $100$ is in the $\ell^{-2}$ regime. 
\Eref{eq:analytic-kurtosis-vs-multipole} also predicts a scaling with the model parameters ($\zeta_0$, $\xi_0$, and $\mathcal{A}$) that agrees well with \fref{fig:kurtosis}.  
For low multipoles, the formula implies $\Delta \hat{\kappa}_\ell \propto 1/\zeta_0$, which is consistent with the numerical results in the figure insofar as lowering $\zeta_0$ increases the excess kurtosis for $\ell \lesssim 30$. 
For high multipoles, the formula implies $\Delta \hat{\kappa}_\ell \propto \zeta_0$, indicating a reversal of the scaling with $\zeta_0$.  
The same reversal is seen on the figure, although the linear $\propto \zeta_0$ scaling is not observed.  
This is possibly because we only show multipoles up to $\ell = 100$, whereas larger values of $\ell$ are required to exhibit the linear scaling.  
Additionally, \eref{eq:analytic-kurtosis-vs-multipole} implies the relation $\Delta\hat{\kappa}_\ell \propto \xi_0^{-1}$, which we have also verified with numerical simulations taking $\mathcal{A} = \zeta_0 = 1$ and $\xi_0 = 0.1$, $1$, and $10$ (results not shown here).  
\Eref{eq:analytic-kurtosis-vs-multipole} implies that $\Delta \hat{\kappa}_\ell$ is independent of $\mathcal{A}$, and this is because $\mathcal{A}$ does not impact the average number of loops $\bar{N}_\ell$; more generally, $\mathcal{A}$ cancels from the kurtosis calculation entirely.  

To conclude, let us address the issues of observability and cosmic variance. 
For a single realization of the CMB sky, one can measure the excess kurtosis using an unbiased kurtosis estimator.  
We consider a simple excess kurtosis estimator defined by 
\begin{align}
    \Delta \hat{\kappa}_\ell^{(1)} = \frac{1}{\ell}\sum_{m = 1}^{\ell} \frac{| \hat{\alpha}_{\ell m} |^4}{(C_\ell^{\alpha\alpha})^2} - 2 
    \ ,
\end{align}
which is motivated by the assumption that the birefringence power spectrum is measured well enough that the true power spectrum $C_\ell^{\alpha\alpha}$ is approximately well known. 
One can apply $\Delta \hat{\kappa}_\ell^{(1)}$ to a measurement of anisotropic CMB birefringence to estimate the excess kurtosis.  
If the moments $\hat{\alpha}_{\ell m}$ were a Gaussian random field, then the mean of this estimator would vanish (Gaussian variables have zero kurtosis), and the standard deviation would be $\mathrm{StDev} \, \Delta \hat{\kappa}_\ell^{(1)} = \sqrt{20 / \ell}$. 
This spread in the estimator, even for Gaussian statistics, is a form of cosmic variance.  
To assess whether the excess kurtosis would be observable for a given model, we can compare the predicted excess kurtosis from \eref{eq:analytic-kurtosis-vs-multipole} with the typical variation $\sqrt{20/\ell}$.  
For the parameters shown in \fref{fig:kurtosis}, the predicted excess kurtosis typically falls below the cosmic variance across a wide range of multipoles.  
On the other hand, in models with small values of $\zeta_0$ and $\xi_0$, the predicted kurtosis can be larger, especially at low multipoles.

\section{Bispectrum}
\label{sec:bispectrum}

A widely-used measure of non-Gaussianity in studies of CMB temperature and polarization anisotropies is the bispectrum, and here we turn our attention to the birefringence bispectrum.  
We denote the first few moments of the multipole moment coefficients $\hat{\alpha}_{\ell m}$ as 
\bsa{}{
    \bar{\alpha}_{\ell_1 m_1} & = \langle \hat{\alpha}_{\ell_1 m_1} \rangle \\ 
    P_{\ell_1 m_1 \ell_2 m_2} & = \langle \hat{\alpha}_{\ell_1 m_1} \hat{\alpha}_{\ell_2 m_2} \rangle \\ 
    B_{\ell_1 m_1 \ell_2 m_2 \ell_3 m_3} & = \langle \hat{\alpha}_{\ell_1 m_1} \hat{\alpha}_{\ell_2 m_2} \hat{\alpha}_{\ell_3 m_3} \rangle 
    \;.
}
For axion-string-induced birefringence, the 1-point functions vanish $\bar{\alpha}_{\ell m} = 0$.  
If the map is statistically isotropic and parity invariant, the 2-point and 3-point functions can be written in terms of the angular power spectrum $C_\ell$ and the reduced bispectrum $b_{\ell_1 \ell_2 \ell_3}$ through the relations~\cite{Komatsu:2001ysk,Planck:2013wtn}
\bsa{}{
    P_{\ell_1 m_1 \ell_2 m_2} & = (-1)^{-m_2} \delta_{\ell_1 \ell_2} \delta_{m_1 -m_2} C_{\ell_1} \\ 
    B_{\ell_1 m_1 \ell_2 m_2 \ell_3 m_3} & = h_{\ell_1 \ell_2 \ell_3}  \begin{pmatrix} \ell_1 & \ell_2 & \ell_3 \\ m_1 & m_2 & m_3 \end{pmatrix} b_{\ell_1 \ell_2 \ell_3} 
    \;,
}
where $h_{\ell_1 \ell_2 \ell_3}$ is a geometrical factor given by
\begin{align}
    h_{\ell_1 \ell_2 \ell_3} = 
    \sqrt{
        \frac{(2\ell_1 + 1)(2\ell_2 + 1)(2\ell_3 + 1)}{4\pi}
    }
    \begin{pmatrix}
        \ell_1 & \ell_2 & \ell_3 \\
        0 & 0 & 0 
    \end{pmatrix}
    \;,
\end{align}
and where the second factor is a Wigner $3$-$j$ symbol.  
The $3$-$j$ symbols vanish unless the multipole moment indices obey the triangle inequality $|\ell_1 - \ell_2| \leq \ell_3 \leq \ell_1 + \ell_2$ (and similarly for the other two index permutations), implying that one can think of $\ell_1$, $\ell_2$, and $\ell_3$ as the lengths of the legs of a triangle. 
Additionally parity invariance requires the bispectrum to vanish unless $\ell_1 + \ell_2 + \ell_3$ is an even integer, and this parity condition is enforced by the geometrical factor $h_{\ell_1 \ell_2 \ell_3}$.  
It is useful to define the random variables~\cite{Komatsu:2001ysk}: 
\bsa{eq:estimators}{
    \hat{C}_{\ell} & =
    (2\ell+1)^{-1} \sum_{m=-\ell}^{\ell} %
    \hat{\alpha}_{\ell m}
    \hat{\alpha}_{\ell m}^\ast 
    \label{eq:estimators-cl}\\ 
    \hat{b}_{\ell_1 \ell_2 \ell_3} & = 
    h_{\ell_1 \ell_2 \ell_3}^{-1}
    \sum_{m_1 = -\ell_1}^{\ell_1} 
    \sum_{m_2 = -\ell_2}^{\ell_2} 
    \sum_{m_3 = -\ell_3}^{\ell_3} 
    \begin{pmatrix}
    \ell_1 & \ell_2 & \ell_3 \\
    m_1 & m_2 & m_3
    \end{pmatrix} 
    \hat{\alpha}_{\ell_1 m_1}
    \hat{\alpha}_{\ell_2 m_2}
    \hat{\alpha}_{\ell_3 m_3} 
    \label{eq:estimators-bispectrum}
    \;,
}
which are unbiased estimators of the angular power spectrum and reduced bispectrum in the sense that $\langle \hat{C}_\ell \rangle = C_\ell$ and $\langle \hat{b}_{\ell_1 \ell_2 \ell_3} \rangle = b_{\ell_1 \ell_2 \ell_3}$.  

The bispectrum is a measure of the non-Gaussianity in the birefringence map.  
This can be understood as follows.  
If the $\hat{\alpha}_{\ell m}$ were independent Gaussian random variables, then higher-point functions could be reduced to 1- and 2-point functions by applying Isserlis's theorem (Wick's theorem).  
Since the 1-point functions vanish, one would expect the 3-point functions to vanish as well implying $b_{\ell_1 \ell_2 \ell_3} = 0$ for a Gaussian birefringence map.  
Conversely, the presence of non-Gaussianity allows the bispectrum to be nonzero, $b_{\ell_1 \ell_2 \ell_3} \neq 0$.  
However, this need not be the case, and it is possible for a non-Gaussian birefringence map to have a vanishing bispectrum $b_{\ell_1 \ell_2 \ell_3} = 0$, and the non-Gaussianity only manifests itself in higher order moments such as the 4-point functions (trispectrum, kurtosis).  
In particular, although axion-string-induced birefringence is non-Gaussian, we nevertheless expect the bispectrum to vanish.  
This is because any configuration of loops that would give rise to a nonzero 3-point function has an equiprobable `opposite' with all loop orientations reversed, which cancels this contribution in the ensemble average. 
However, it's important to bear in mind that although the bispectrum may vanish as an ensemble average $b_{\ell_1 \ell_2 \ell_3} = 0$, its estimator must be nonzero for any given realization $\hat{b}_{\ell_1 \ell_2 \ell_3} \neq 0$.  
Here we are primarily interested in evaluating the typical size of the bispectrum estimator, quantified through its standard deviation $\mathrm{StDev}[\hat{b}_{\ell_1 \ell_2 \ell_3}]$. 

\begin{figure}[t!]
    \centering
    \begin{minipage}{6in}
      \centering
      \raisebox{-0.5\height}{\includegraphics[width=0.45\textwidth]{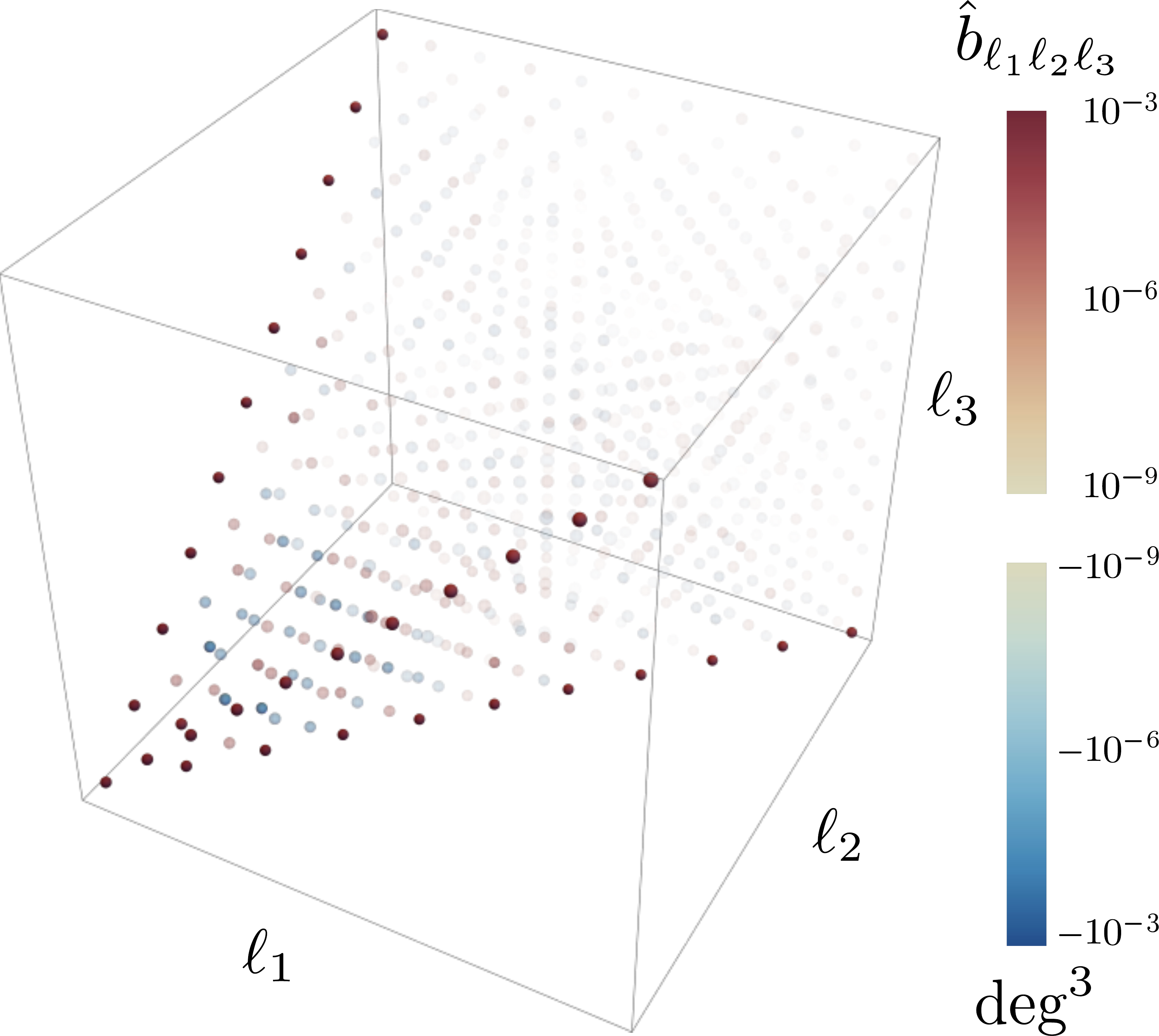}}
      \hspace*{.05in}
      \raisebox{-0.545\height}{\includegraphics{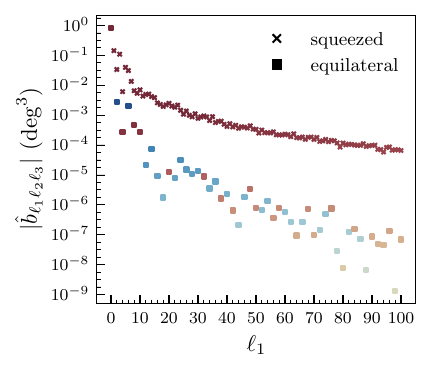}}
    \end{minipage}
    \caption{\label{fig:bispectrum_realization}
    A single realization of the bispectrum estimator $\hat{b}_{\ell_1 \ell_2 \ell_3}$ calculated from a simulated birefringence map using the loop-crossing model with parameters $\zeta_0 = \xi_0 = \mathcal{A} = 1$.  For other values of $\mathcal{A}$ the bispectrum estimator would scale as $\propto \mathcal{A}^3$.  \textit{Left:}  Colored dots indicate values of the bispectrum estimator for multipole moment indices $\ell_1, \ell_2, \ell_3$ ranging from $0$ to $100$ in steps of $10$.  \textit{Right:} Values of the bispectrum estimator along the edge of the tetrahedron where $\ell_1 = \ell_2$ and $\ell_3 = 0$ corresponding to a `squeezed' triangle (cross markers) and along the main diagonal where $\ell_1 = \ell_2 = \ell_3$ corresponding to an `equilateral' triangle (square markers). 
    }
\end{figure}

To assess the typical bispectrum arising from axion-string-induced birefringence, we have used the loop-crossing model to simulate a single realization of the birefingence map and calculate the bispectrum estimator $\hat{b}_{\ell_1 \ell_2 \ell_3}$.  
These results are presented in \fref{fig:bispectrum_realization}.  
On the left we show a visualization of $\hat{b}_{\ell_1 \ell_2 \ell_3}$ where the multipole moment indices $(\ell_1, \ell_2, \ell_3)$ are mapped to points in a three-dimensional volume.  
Colored dots indicate the value of the bispectrum estimator on a log scale, and smaller values are rendered as semi-transparent to enhance visibility.  
The tetrahedral shape is a consequence of the triangle inequalities ($|\ell_1 - \ell_2| \leq \ell_3 \leq \ell_1 + \ell_2$ and permutations), since the bispectrum estimator vanishes outside of this region due to geometrical constraints imposed by the $3$-$j$ symbols. 
Additionally the parity condition requires $\ell_1 + \ell_2 + \ell_3$ to be an even integer, which further causes many $\hat{b}_{\ell_1 \ell_2 \ell_3}$ to vanish. 
The right panel plots the bispectrum estimator along two rays through the tetrahedron. 
These rays correspond to (1) the main diagonal of the tetrahedron along which $\ell_1 = \ell_2 = \ell_3$, corresponding to the equilateral triangle form; and (2) the edge of the tetrahedron along which $\ell_1 = \ell_2$ and $\ell_3 = 0$, corresponding to the squeezed triangle form. 
Due to the symmetry properties of the $3$-$j$ symbols, the values of $\hat{b}_{\ell_1 \ell_2 \ell_3}$ along the three tetrahedral edges are identical. 

Several qualitative features of \fref{fig:bispectrum_realization} are easily understood.  
Since the bispectrum $b_{\ell_1 \ell_2 \ell_3}$ is expected to vanish for axion-string-induced birefringence, it is not surprising to see that the bispectrum estimator $\hat{b}_{\ell_1 \ell_2 \ell_3}$ evaluates to a scatter of positive and negative values.  
For $\ell_1 = \ell_2 = \ell_3 = 0$ the bispectrum estimator is simply the cube of the monopole multipole moment coefficient $\hat{b}_{000} = \sqrt{4\pi} (\hat{\alpha}_{00})^3$, and using $\hat{\alpha}_{00} \sim 0.5 \, \mathrm{deg}$ from \fref{fig:alpha_lm_distributions} (same simulation parameters) gives $\hat{b}_{000} \sim 0.4 \, \mathrm{deg}^3$, which is compatible with the figure.  
Moving to larger $\ell$, the bispectrum estimator tends to decrease in magnitude for higher multipoles, and we quantify and discuss this behavior further below.  
For this realization the bispectrum estimator is positive along the three tetrahedral edges, corresponding to the squeezed triangle form, but for other realizations they may be negative.  
The sign of $\hat{b}_{\ell_1 \ell_2 \ell_3}$ along these rays are correlated with the random sign of the monopole $\hat{\alpha}_{00}$.  
One can prove this using identities of the Wigner $3$-$j$ symbols, but heuristically the relation is $\hat{b}_{\ell \ell 0} \sim \langle |\hat{\alpha}_{\ell m}|^2 \hat{\alpha}_{00} \rangle$.  

\begin{figure}[t!]
      \centering
      \includegraphics{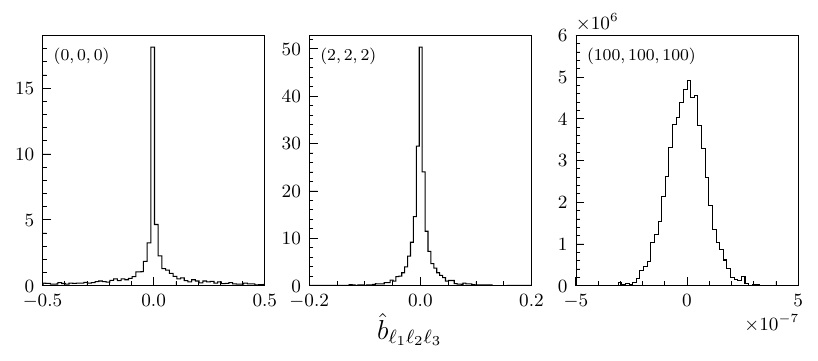}
      \caption{\label{fig:bispectrum_ensemble}
      Distributions of bispectrum estimators for $\ell_1 = \ell_2 = \ell_3 = 0$ (left), $2$ (middle), and $100$ (right). We have used $5,\!000$ simulations of the loop-crossing model with parameters $\zeta_0 = \xi_0 = \mathcal{A} = 1$.  For other values of $\mathcal{A}$ the bispectrum estimator would scale as $\propto \mathcal{A}^3$.
      }
\end{figure}

Repeating these simulations $5,\!000$ times with the same LCM model parameters ($\zeta_0 = \xi_0 = \mathcal{A} = 1$), we evalaute the bispectrum estimator for each realization and present a sample of these results in \fref{fig:bispectrum_ensemble}.  
We show histograms over the bispectrum estimator for $\ell_1 = \ell_2 = \ell_3 = 0$, $2$, and $100$, which are normalized so that their integral equals $1$. 
These distributions appear to be centered at $\hat{b}_{\ell_1 \ell_2 \ell_3} = 0$, and they are approximately symmetric. 
Moreover, we have verified that the sample mean falls like $1 / \sqrt{N_{\rm sims}}$, as one expects for a random variable with vanishing mean.
The distributions in \fref{fig:bispectrum_ensemble} appear visibly non-Gaussian for $\ell = 0$ and $2$, but this is not evidence of non-Gaussianity, since the product $\hat{b}_{\ell_1 \ell_2 \ell_3} \sim \hat{\alpha}_{\ell_1 \ell_2 \ell_3}^3$ would be non-Gaussian even if the individual factors $\hat{\alpha}_{\ell_1 \ell_2 \ell_3}$ were Gaussian.  
For $\ell = 100$ the distribution appears Gaussian, and this can be understood from the central limit theorem: since the bispectrum estimator is a sum over many terms $\hat{b}_{\ell_1 \ell_2 \ell_3} \sim \sum \hat{\alpha}^3$, see \eref{eq:estimators}, we expect that $\hat{b}_{\ell\ell\ell}$ should be approximately normally distributed at high $\ell$ since $\hat{b}_{\ell\ell\ell}$ is a linear combination of many i.i.d. random variables.
The width of the histogram decreases for increasing multipole moment index $\ell$, which is compatible with the trend seen already in \fref{fig:bispectrum_realization}.  

Although the bispectrum vanishes upon ensemble averaging, it is nonzero for each realization. 
Such fluctuations could still impact CMB polarization data, where only one realization is available. 
This observation motivates us to evaluate the standard deviation of the bispectrum estimator $\mathrm{StDev}[\hat{b}_{\ell_1 \ell_2 \ell_3}] = [\langle \hat{b}_{\ell_1 \ell_2 \ell_3}^2 \rangle - \langle \hat{b}_{\ell_1 \ell_2 \ell_3} \rangle^2]^{1/2}$. 
If the birefringence map were Gaussian, the 6-point function $\langle \hat{b}_{\ell_1 \ell_2 \ell_3}^2 \rangle \sim \langle \hat{\alpha}_{\ell m}^6 \rangle$ could be reduced to products of 2-point functions using Isserlis's theorem.  
By doing so we find 
\bes{\label{eq:sigma_l1_l2_l3_Gaussian}
    & {\rm StDev}\big[ \hat{b}_{\ell_1 \ell_2 \ell_3} \big]_\text{if $\hat{\alpha}_{\ell m}$ are Gaussian}
    =
    | h_{\ell_1 \ell_2 \ell_3} |^{-1}
    \sqrt{C_{\ell_1} C_{\ell_2} C_{\ell_3}}
    \\ & \quad 
    \times \Bigl[ 
    1
    + 2 \delta_{\ell_1 \ell_2} \delta_{\ell_2 \ell_3} 
    + \delta_{\ell_2 \ell_3} + \delta_{\ell_1 \ell_2} 
    + \delta_{\ell_3 \ell_1} 
    + 6 \, \delta_{\ell_1 0} \, \delta_{\ell_2 0} \, \delta_{\ell_3 0} 
    \\ & \quad \qquad 
    + (2\ell_1 + 1) \, \delta_{\ell_1 \ell_2}\, \delta_{\ell_3 0} 
    + (2\ell_2 + 1) \, \delta_{\ell_2 \ell_3}\, \delta_{\ell_1 0} 
    + (2\ell_3 + 1) \, \delta_{\ell_3 \ell_1}\, \delta_{\ell_3 0}
    \Big]^{1/2}
    \;,
}
where $C_\ell$ is the angular power spectrum, and we assumed that the multipole indices obey the triangle inequality and parity condition; variations of this formula (bispectrum covariance) appear in refs.~\cite{Luo:1993xx,Spergel:1999xn,Gangui:1999vg,Gangui:2000gf}. 
For a scale-invariant power spectrum $\ell(\ell+1) C_\ell$ is independent of $\ell$, and one expects to find $\mathrm{StDev}[ \hat{b}_{\ell_1 \ell_2 \ell_3}] \propto \ell^{-7/2}$ in the equilateral configuration and a larger $\ell^{-2}$ in the squeezed configuration.  
We are interested in whether departures from this scaling can arise from the inherent non-Gaussianity of axion-string-induced birefringence.  

\begin{figure}[t!]
      \centering
      \includegraphics{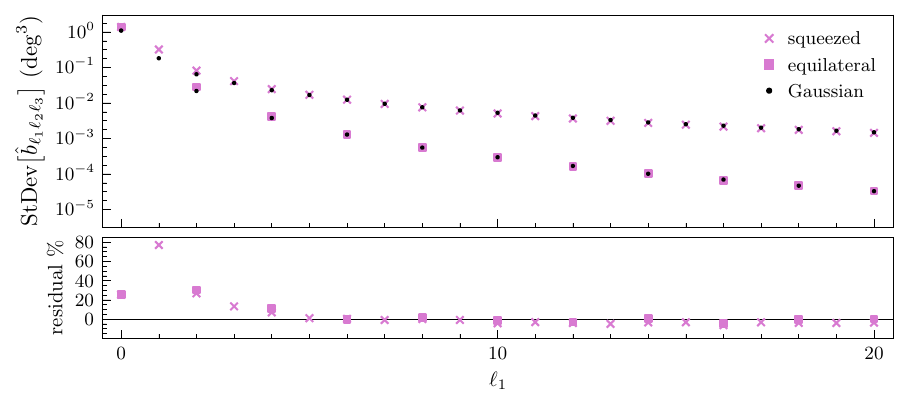}
      \caption{\label{fig:bispectrum-stdev}
      \textit{Top:} Standard deviation of the birefringence bispectrum estimator ${\rm StDev}[\hat{b}_{\ell_1 \ell_2 \ell_3}]$. We show all values that are nonzero for $\ell \leq 20$; no binning was performed. Purple markers denote results from axion-string-induced birefringence, calculated as the sample average of $5,\!000$ LCM simulations for the model with parameters $\zeta_0 = \xi_0 = \mathcal{A} = 1$. Crossed markers correspond to the squeezed triangle form with $\ell_1 = \ell_2,\ \ell_3 = 0$, and square markers correspond to the equilateral triangle with $\ell_1 = \ell_2 = \ell_3$.  Black dots indicate the expected bispectrum standard deviation for Gaussian birefringence, given by \eref{eq:sigma_l1_l2_l3_Gaussian}. \textit{Bottom:} The fractional difference between the bispectrum standard deviation and the expectation for Gaussian statistics.  
      }
\end{figure}

In \fref{fig:bispectrum-stdev} we show the sample standard deviation of the bispectrum estimator $\mathrm{StDev}[\hat{b}_{\ell_1 \ell_2 \ell_3}]$, calculated using the same loop-crossing model parameters as in the previous figure, $\zeta_0 = \xi_0 = \mathcal{A} = 1$.  
The purple crosses and boxes correspond to axion-string-induced birefringence, and they were calculated using our simulations; the black dots correspond to Gaussian birefringence, and they were calculated using \eref{eq:sigma_l1_l2_l3_Gaussian}.  
To evaluate $C_\ell$ in \eqref{eq:sigma_l1_l2_l3_Gaussian} we performed $1,\!000$ LCM simulations and averaged over the power spectrum estimator $\hat{C}_\ell$ for each realization, which is approximately scale invariant for $\ell \lesssim 100$.  
From the figure, it can be seen that the standard deviation of the bispectrum tends to track closely to the expectation for Gaussian birefringence, particularly at higher multipoles with $\ell \gtrsim 5$.  
This explains why the bispectrum tends to be larger for the squeezed configuration as compared with the equilateral configuration, and why they decrease toward larger $\ell$ while approximately tracking power laws. 
For low multipoles $\ell \lesssim 4$ the difference between the bispectrum standard deviation and the Gaussian expectation can be significant, reaching a maximum fractional difference of approximately $80\%$ for $\ell_1 = \ell_2 = 1$ and $\ell_3 = 0$.  
Since the bispectrum tends to be larger than the Gaussian expectation across a range of low multipoles, correlated measurements could be used to search for evidence of non-Gaussian axion-string-induced birefringence.  

\section{Conclusion}
\label{sec:conclusion}

If a network of axion strings is present in the Universe after recombination, then a coupling of the axion-like particles to electromagnetism will induce anisotropic cosmic birefringence.  
The birefringence angle will vary across the sky tracing the `shadow' of the cosmic string network with sharp edges and loop-like features.  
The statistics of this birefringence map are therefore non-Gaussian, since a Gaussian random field would resemble featureless noise.  
In this work we have used two familiar measures of non-Gaussianity, kurtosis and bispectrum, to quantify the departure from Gaussian statistics.  

Kurtosis is calculated from the fourth moment of the birefringence rotation angle; roughly $\kappa \sim \langle \hat{\alpha}^4 \rangle / \langle \hat{\alpha}^2 \rangle^2$.  
For Gaussian statistics, four-point functions are equal to products of two-point functions, and we define the excess kurtosis $\Delta \hat{\kappa}_\ell$ to measure the deviation from Gaussian statistics.  
We use a combination of numerical simulation, in a phenomenological framework called the loop-crossing model, and analytical approximation to evaluate the excess kurtosis across a range of angular scales (with multiple index $\ell = 0$ to $100$) and for a range of string network model parameters ($\zeta_0$, $\xi_0$, and $\mathcal{A}$).  
We find that excess kurtosis tends to be positive, order $0.1$ to $10$ at the largest angular scales (depending on model parameters), and decreasing toward smaller angular scales.  
To understand how the excess kurtosis varies with $\ell$ and depends on the model parameters, we have developed a simplified analytical model that leads to the approximation for $\Delta \hat{\kappa}_\ell$ provided in \eref{eq:analytic-kurtosis-vs-multipole}.  
This formula agrees remarkably well with the scaling relations inferred from simulations.  
To assess observability, we have calculated the cosmic variance of an excess kurtosis estimator assuming $\hat{\alpha}$ to be a Gaussian random field and perfect knowledge of the power spectrum $C_\ell^{\alpha\alpha}$.
For small values of $\zeta_0$ and $\xi_0$, the excess kurtosis arising from axion-string induced birefringence can be larger, on average, than the uncertainty from cosmic variance.  
These estimates indicate that the signal is detectable in principle, but likely challenging in practice.  

The bispectrum is defined as the third moment of the birefringence rotation angle at different angular scales; roughly $b \sim \langle \hat{\alpha}_1 \hat{\alpha}_2 \hat{\alpha}_3 \rangle$.  
For axion-string-induced birefringence we expect the bispectrum to vanish as an ensemble average, but it must be nonzero in any given realization, and our analysis focuses on calculating its standard deviation.  
Using numerical simulations of the loop-crossing model, we evaluate the reduced bispectrum $\hat{b}_{\ell_1 \ell_2 \ell_3}$ for a range of angular scales from $\ell_i = 0$ to $100$.  
We find that the bispectrum tends to be largest for the `squeezed' triangle form ($\ell_1 = 0$, $\ell_2 = \ell_3$ and permutations) and relatively smaller in the `equilateral' triangle form ($\ell_1 = \ell_2 = \ell_3$).  
For both cases the typical bispectrum decreases toward larger $\ell_i$, approximately tracking a power law.  
We discuss how these general trends would arise even if the birefringence rotation angle followed Gaussian statistics.  
For the model parameters that we explored numerically here ($\zeta_0 = \xi_0 = \mathcal{A} = 1$), the typical bispectrum tracks the Gaussian expectation, and the largest difference occurs for $\ell_1 = \ell_2 = 1$ and $\ell_3 = 0$ (and permutations) where the fractional difference is approximately $80\%$.  
This deviation suggests that an anomalously large bispectrum would be consistent with axion-strings, although additional information such as a measurement of the power spectrum would be needed to claim evidence of axion strings from CMB polarization.  

The work presented here serves to better characterize the cosmological signatures of an axion string network present in the Universe after recombination.  
If evidence for anisotropic birefringence is detected in CMB polarization measurements using two-point statistics, such as $EB$ cross-correlation, the higher-moment statistics studied here will prove valuable to discriminate across different possible new physics sources of birefringence. 
For instance, at the level of the power spectrum the parameters of axion-string-induced birefringence exhibit a degeneracy; the signal is proportional to $\mathcal{A}^2 \xi_0$ where the anomaly coefficient $\mathcal{A}$ quantifies the strength of the axion-photon coupling, and the loop density parameter $\xi_0$ controls the number of axion string loops per Hubble volume.  
A detection of anisotropic birefringence and a measurement of its power spectrum would not provide sufficient information to discriminate between $\mathcal{A}$ and $\xi_0$.  
However, in general this degeneracy can be broken by higher-point statistics \cite{Yin:2023vit}.  
For example, \eref{eq:analytic-kurtosis-vs-multipole} reveals that the excess kurtosis $\Delta \hat{\kappa}_\ell$ is insensitive to $\mathcal{A}$ and goes inversely with $\xi_0$.  
Consequently, with sufficient information it becomes possible to independently determine the properties of the axion string network, parametrized here by $\zeta_0$ and $\xi_0$, and the fundamental parameters of the new physics, parametrized by the anomaly coefficient $\mathcal{A}$ as well as the axion mass $m_a$.  

\acknowledgments
We are grateful to Mustafa A. Amin, Mudit Jain, and Joel Meyers for discussions and comments on the draft.  
We thank Winston Yin for pointing out an incorrect reference in the first version of this paper.  
R.H. is grateful to Siyang Ling for many insightful conversations.
R.H.~and A.J.L.~are supported in part by the National Science Foundation under Award No.~PHY-2114024. 
Some of the results in this paper have been derived using the healpy and HEALPix package. 

\appendix

\section{Analytical analysis for kurtosis}
\label{app:kurtosis}

To develop an analytical understanding of the kurtosis arising in axion-string-induced birefringence, we provide here a simplified description that is analytically tractable.  
We first consider the monopole $\hat{\alpha}_{00}$ and then extend this analysis to higher multipoles with $\ell > 0$.  

\subsection{Monopole}

Consider the monopole of the birefringence map
\ba{\label{eq:app-monopole}
    \hat{\alpha}_{00} 
    = \int \! \dd^2 n \, Y^*_{00}(\nhat)\,\hat{\alpha}(\nhat) 
    = \frac{1}{\sqrt{4\pi}} \int \! \dd^2 n \, \hat{\alpha}(\nhat) 
    \;.
}
In the loop-crossing model, the birefringence map $\hat{\alpha}(\nhat)$ is built up from random overlapping string loops of different sizes and orientations, distributed isotropically across the sky.  
Photons propagating through the disk encircled by a loop experience a random birefringence, which accumulates with multiple loop crossings.
For simplicity we suppose here that every loop crossing leads to a statistically equivalent shift in the monopole, $\Delta \hat{\alpha}_{00} = +C$ or $-C$ with equal probability.  
More realistically in the loop-crossing model, larger loops contribute more and smaller loops less, and the loop's orientation affects the solid angle it spans on the sky, but these effects are ignored for this simplified analysis.  
Note that the location of the loops on the sky is irrelevant for the monopole.  
We also suppose that the number of loops giving this contribution, denoted as $\hat{N}_\mathrm{loops}$ is random and Poisson distributed with intensity parameter $\bar{N}_\mathrm{loops}$.  
The quantity $\bar{N}_\mathrm{loops}$ is calculable within the loop-crossing model in terms of the properties of the string network.  
These simplifications allow the monopole to be written as 
\bes{
    \hat{\alpha}_{00} & = C \sum_{i=1}^{\hat{N}_\mathrm{loops}} \hat{W}_i \\ 
    \hat{W}_i & \sim - \text{1 or 1 with equal probability} \\ 
    \hat{N}_\mathrm{loops} & \sim {\rm Poisson}(\bar{N}_\mathrm{loops}) 
    \;.
}
This is an example of a hierarchical random model, where the number of random variables (loop crossings) is itself a random variable.  

We are interested in the moments of $\hat{\alpha}_{00}$, which give the kurtosis.  
It is useful to recognize that $(\hat{W}_i + 1) / 2$ is a ${\rm Bernoulli}(1/2)$ random variable, taking values $0$ and $1$ with equal probability.  
The sum over a sequence of $n$ i.i.d. ${\rm Bernoulli}(1/2)$ random variables is a ${\rm binomial}(n,\, 1/2)$ random variable. 
This motivates us to define $\hat{Y}_n = \sum_{i=1}^n (\hat{W}_i + 1)/2$ and write the monopole as 
\ba{\label{eq:monopole-hierarchy}
    \hat{\alpha}_{00} & = C \Bigl( 2 \hat{Y}_{\hat{N}_\mathrm{loops}} - \hat{N}_\mathrm{loops} \Bigr) 
    \;.
}
This expression can be used to calculate the kurtosis of $\hat{\alpha}_{00}$ analytically using the fact that for any two random variables $\hat{x}$ and $\hat{y}$, expectation values can be calculated as \cite{CaseBerg:01} 
\ba{
    \mathrm{E}(\hat{x}) = \mathrm{E}\bigl[ \mathrm{E}(\hat{x} | \hat{y}) \bigr] 
}
so long as the expectation values exist. 
For example, the first moment is calculated as follows: 
\bes{
    \frac{1}{C} \, \mathrm{E}(\hat{\alpha}_{00})
    & = 2 \, \mathrm{E}\bigl[ \hat{Y}_{\hatNloops} \bigr] - \mathrm{E}(\hatNloops) \\ 
    &= 2 \, \mathrm{E}\bigl[ \mathrm{E}(\hat{Y}_{\hatNloops}|\hatNloops) \bigr] - \bar{N}_\mathrm{loops} \\
    & = 2 \, \mathrm{E}\bigl[\hatNloops / 2 \bigr] - \bar{N}_\mathrm{loops} \\ 
    & = 0
    \;.
}
Repeating this procedure for $\mathrm{E}(\hat{\alpha}_{00}^2)$, and $E(\hat{\alpha}_{00}^4)$ we find
\begin{align}
    \mathrm{E}(\hat{\alpha}_{00}^2) & = C^2 \, \bar{N}_\mathrm{loops} \\
    \mathrm{E}(\hat{\alpha}_{00}^4) & = C^4 \, \Bigl[ 3 \bar{N}_\mathrm{loops}^2 + \bar{N}_\mathrm{loops} \Bigr] 
    \;.
\end{align}
The corresponding excess kurtosis is
\begin{align}\label{eq:monpole_excess_kurtosis}
    \Delta \kappa_{0} 
    = \kappa_{00} - 3 
    = \frac{\mathrm{E}(\hat{a}_{00}^4)}{\mathrm{E}(\hat{a}_{00}^2)^2} - 3 
    = 1 / \bar{N}_\mathrm{loops} 
    \;,
\end{align}
which is the result quoted in the main text. 

\subsection{Higher multipoles}

We suppose that the monopole relation in \eref{eq:monpole_excess_kurtosis} extends to higher multipoles as 
\ba{
    \Delta \kappa_\ell \sim 1 / \bar{N}_\ell 
    \;,
}
where $\bar{N}_\ell$ denotes the average number of loops on an angular scale $\sim \pi / \ell$.  
By calculating $\bar{N}_\ell$ in the loop-crossing model, we obtain an expression for the angle-averaged excess kurtosis $\Delta \kappa_\ell$ in terms of the multipole index $\ell$ and the string network model parameters.  

First, in the loop-crossing model, the typical length of loops in the network grows with time to track the growing Hubble scale.  
At redshift $z$, the typical angular scale of the loops is \cite{Jain:2021shf} 
\ba{
    \delta \theta 
    \sim \pi / \ell 
    \sim 
    \frac{2 \lambda \zeta_0}{a(z) H(z) s(z)} 
    \;,
}
where $\lambda = 0.3$ accounts for the random orientation of the loops, and in a matter-dominated cosmology: $a(z) \propto (1+z)^{-1}$ is the scale factor, $H(z) \propto (1+z)^{-3/2}$ is the Hubble parameter, and $s(z) = \int_0^z \! \mathrm{d}z^\prime / a_0 H(z^\prime)$ is the comoving distance to redshift $z$. 
Solving this relation for $z$ gives 
\ba{
    z_\ell \sim \frac{\lambda \zeta_0 \ell (2\pi + \lambda \zeta_0 \ell)}{\pi^2} 
    \;,
}
which represents the redshift at which loops with angular scale $\pi/\ell$ were present in the network. 
In the loop-crossing model, the average comoving number density of loops in the network at redshift $z$ is taken to be 
\ba{
    \bar{n}(z) 
    = \frac{\xi_0 a(z)^3 H(z)^3}{2\pi\zeta_0} 
    \;.
}
Integrating over a spherical shell of redshifts $z_\ell < z < z_\ell + \Delta z$ gives 
\bes{\label{eq:num_loops}
    \bar{N}_\ell 
    & = \int_{z_\ell}^{z_\ell+\Delta z} \! \dd z \, 4\pi s^2(z) \frac{\dd s}{\dd z} \, \bar{n}(z) \\ 
    & \approx \Delta z \, 4\pi s^2(z_\ell) \frac{1}{a_0 H(z_\ell)} \, \frac{\xi_0 a(z_\ell)^3 H(z_\ell)^3}{2\pi\zeta_0} \\ 
    & \sim \frac{8 \lambda^2 \zeta_0 \xi_0 \ell^2 \Delta z}{(\pi + \lambda \zeta_0 \ell)^2} 
    \;,
}
which represents the average number of loops with angular extent $\pi/\ell$.  
If the excess angle-averaged kurtosis can be estimated as $\Delta \kappa_\ell \sim 1 / \bar{N}_\ell$, then we have 
\ba{
    \Delta \kappa_\ell \sim \frac{\zeta_0}{8 \, \xi_0} \biggl( 1 + \frac{\pi}{\lambda \zeta_0 \ell} \biggr)^2
}
where we have taken $\Delta z = 1$.  
This expression matches the result quoted in the main text.  

\bibliographystyle{JHEP}
\bibliography{refs}

\end{document}